\documentclass[useAMS,usenatbib]{mn2e}

\usepackage{graphicx}
\usepackage{amsmath}
\usepackage{indentfirst}
\def\kms {$\rm km\,s^{-1} \,$}

\title[Gas excitation and kinematics in Pictor\,A]{Integral Field Spectroscopy of the circumnuclear region of the Radio Galaxy Pictor A}

\author[Guilherme S. Couto et al.]{Guilherme S. Couto$^{1}$\thanks{E-mail:gcouto@if.ufrgs.br}, Thaisa Storchi-Bergmann$^{1}$, Andrew Robinson$^{2}$,
\newauthor Rogemar A. Riffel$^{3}$, Preeti Kharb$^{4}$, Davide Lena$^{2,5,6}$ and Allan Schnorr-M{\"u}ller$^{7}$\\
$^{1}$Universidade Federal do Rio Grande do Sul, IF, CP 15051, Porto Alegre 91501-970, RS, Brazil\\
$^{2}$School of Physics and Astronomy, Rochester Institute of Technology, 85 Lomb Memorial Dr., Rochester, NY 14623, USA\\
$^{3}$Universidade Federal de Santa Maria, Departamento de F\'isica, Centro de Ci\^encias Naturais e Exatas, 97105-900, Santa Maria, RS, Brazil\\
$^{4}$Indian Institute of Astrophysics, 2nd Block, Koramangala, Bangalore 560034, India\\
$^{5}$SRON Netherlands Institute for Space Research, Sorbonnelaan 2, NL-3584 CA Utrecht, The Netherlands\\
$^{6}$Department of Astrophysics/IMAPP, Radboud University, Nijmegen, PO Box 9010, NL-6500 GL Nijmegen, The Netherlands\\
$^{7}$Max-Planck-Institute f{\"u}r Extraterrestrische Physik, Postfach 1312, D-85741 Garching, Germany\\}

\begin{document}

\date{Accepted 2016 February 18. Received 2016 February 17; in original form 2015 August 17.}

\pagerange{\pageref{firstpage}--\pageref{lastpage}} \pubyear{2016}

\maketitle

\label{firstpage}

\begin{abstract}
We present optical integral field spectroscopy of the inner $2.5 \times 3.4\,$kpc$^2$ of the broad-line radio galaxy Pictor A, at a spatial resolution of $\approx 400\,$pc. Line emission is observed over the whole field-of-view, being strongest at the nucleus and in an elongated linear feature (ELF) crossing the nucleus from the south-west to the north-east along PA $\approx 70^\circ$. Although the broad double-peaked H$\alpha$ line and the [O\,{\sc i}]6300/H$\alpha$ and [S\,{\sc ii}]6717+31/H$\alpha$ ratios are typical of AGNs, the [N\,{\sc ii}]6584/H$\alpha$ ratio ($0.15-0.25$) is unusually low. We suggest that this is due to the unusually low metallicity of the gas. Centroid velocity maps show mostly blueshifts to the south and redshifts to the north of the nucleus, but the velocity field is not well fitted by a rotation model. Velocity dispersions are low ($< 100\,$\kms) along the ELF, ruling out a jet-cloud interaction as the origin of this structure. The ELF shows both blueshifts and redshifts in channel maps, suggesting that it is close to the plane of the sky. The ELF is evidently photoionized by the AGN, but its kinematics and inferred low metallicity suggest that this structure may have originated in a past merger event with another galaxy. We suggest that the gas acquired in this interaction may be feeding the ELF.
\end{abstract}

\begin{keywords}
Galaxies: individual Pictor\,A -- Galaxies: active -- Galaxies: nuclei -- Galaxies: kinematics -- Galaxies: jets 
\end{keywords}

\section{Introduction}

Nuclear activity in radio galaxies, via ejection of radio plasma, powered by accretion onto their central engines, is known to influence the conditions of the surrounding interstellar medium (ISM), affecting its excitation and ionization. This has been observed in several galaxies, on small and large scales \citep{nesvadba09,tremblay09,rosario10a,couto13,reyfei13,santoro14}. Theoretical studies are in agreement with this scenario, suggesting that the radio activity influences the energetics and thermodynamics of the hot, space-filling medium in these hosts \citep{reynolds02,fragile04,wagbic11}. Radio jet interactions with circumnuclear gas can generate shock induced ionization, in addition to possible photoionization from the AGN \citep{bauhec89,best00}. Such interactions can even produce massive gas outflows, with entrained gas being accelerated to high velocities \citep{feruglio10,morganti13}. The role of such AGN feedback in the evolution of these galaxies, however, is still a subject of ongoing debate. Studies have indicated that shocks resulting from jet interactions can trigger star formation \citep{oosmor05}. On the other hand, AGN feedback may be a key factor in quenching star formation, mainly in late stages of the host galaxy's evolution \citep{dimatteo05,silverman08}. The tight correlations found between the mass of the supermassive black hole (SMBH) and the velocity dispersion (and other properties) of the bulge of the host galaxy, suggest that coevolution of the SMBH and bulge, with AGN feedback playing a major role \citep{fermer00,gebhardt00,korho13}. Studies of thetudies of the ionized nuclear gas in radio galaxies are important to the understanding of the characteristics and effects of such interactions.

Among the objects where signatures of jet-ISM interactions can be found, Pictor A, a famous nearby ($z = 0.035$) radio-bright ($P_{408\, \textrm{MHz}} = 6.7 \times 10^{26} \textrm{W Hz}^{-1}$) Fanaroff-Riley type II galaxy \citep[FRII;][]{fanril74}, is an excellent laboratory in which we can study the circumnuclear gas emission surrounding a powerful radio jet.

For $H_0 = 70.5\,$\kms Mpc$^{-1}$, the galaxy has a luminosity distance of $152.9\,$Mpc, and $1$\arcsec\, corresponds to 690 pc. Its prominent double-lobed radio emission is characteristic of this class and has been extensively studied \citep{perley97,tingay00,tingay08}. The radio jet, which reaches distances $> 200\,$kpc, presents a small bend at about $180\arcsec$ ($\approx 125\,$kpc) from the nucleus, before it reaches the western lobe. This permits different measurement criteria, introducing some uncertainty in the position angle (PA) of the kiloparsec-scale radio jet; however almost all works have reported values of PA $= -75 \pm 5^\circ$ \citep{perley97,simkin99,tingay00,marshall10}. The X-ray jet observed by \citet{wilson01} has a width of $\approx 2\,$kpc and coincides with the radio jet. The radio and X-ray jets are located on the west side of the galaxy, with the X-ray jet being up to $\approx 15\,$ times brighter than any counterjet presumably due to relativistic boosting, indicating that the west 
side is probably the closer side. 

Pictor A is also classified as a broad-line radio galaxy (BLRG), due to its highly broadened ($> 10,000\,$\kms) double-peaked Balmer lines in the optical spectra \citep{halera94,lewis10}. \citet{erahal03} have modeled the double-peaked H$\alpha$ profile as arising in the outer parts of an accretion disk, finding inclinations of $i \approx 30^{\circ}$ and $i > 65^{\circ}$ for elliptical and circular disk models, respectively. The host galaxy morphology is not clear in the optical, but previous studies indicate the presence of a disk. \citet{lauberts82} classifies this object as an Sa, while \citet{loveday96} found no evidence of spiral arms, classifying it as an S0. \citet{inskip10}, using near-IR data, concluded that Pictor A's surface brightness distribution is best modelled by a disky galaxy profile with a nuclear point source. A tidal tail was recently discovered by \citet{gentry15}, starting 18\arcsec\, (12 kpc) north of Pictor A, sweeping to the west along a path about 60 kpc long. This long tail indicates the occurrence of a merger event some hundred million years ago.

In this work, we present a two-dimensional analysis of the kinematics and excitation of the gas within the inner $1.5$ kiloparsec (in radius) of Pictor A. This paper is organized as follows: in Sec. \ref{obs} we describe the observations and data reduction, in Sec. \ref{res} we present the results for the gas excitation and kinematics, in Sec. \ref{disc} we discuss our results and in Sec. \ref{conc} we present our conclusions.

\section{Observations and Data Reduction}
\label{obs}

Two-dimensional optical spectroscopic data were obtained at the Gemini South Telescope with the {\it Gemini Multi-Object Spectrograph Integral Field Unit} \citep[GMOS IFU,][]{allington02} on November 21, 2006, as part of the program GS-2004B-Q-25. The observations comprise nine individual exposures of 600 s centred at $\lambda$\,7766\,\AA\ with a spectral coverage from $\lambda$5600\,\AA\ to $\lambda$9925\,\AA. The R400+\_G5325 grating with the IFU-R mask was used. The spectral resolution is 2.4\,\AA\ at $\lambda$\,7000\,\AA\ ($\approx$\,103 \kms) -- as derived from the full width at half maximum (FWHM) of the CuAr lamp emission lines. The angular resolution -- as determined from the FWHM of the spatial profile obtained along the double-peaked H$\alpha$ broad line profile -- showed an elongation, with a value of $0\farcs57$ along the x-axis and $0\farcs82$ along the y-axis. The GMOS IFU has a rectangular field of view (FOV), of approximately 3$\farcs$6\,$\times$\,4$\farcs$9, corresponding to 2.5\,kpc$\times$\
,3.4\,kpc at the galaxy. The major axis of the IFU was oriented along a PA=$-20^\circ$.

The data reduction was accomplished using tasks in the {\sc gemini.gmos iraf} package as well as generic {\sc iraf}\footnote{IRAF is distributed by the National Optical Astronomy Observatory, which is operated by the Association of Universities for Research in Astronomy (AURA) under a cooperative agreement with the National Science Foundation.} tasks. The reduction procedure included trimming of the images, bias subtraction, flat-fielding, wavelength calibration, sky subtraction and flux calibration. A correction for telluric absorption lines was also applied, since this affected the sulfur emission lines. Also, we have corrected the data for differential atmospheric refraction \citep{steiner09}. Finally, we managed to improve the asymmetric point spread function by applying a Richardson-Lucy deconvolution algorithm \citep{richardson72,lucy74} to the datacube. This resulted in an angular resolution of $0\farcs43$ along the x-axis and $0\farcs56$ along the y-axis (297\,pc$\times$\,386\,pc). A comparison of the original datacube with that resulting from the deconvolution shows that no significant spurious features were introduced. The final IFU data cube contains $\sim 2000\,$ spectra, each spectrum corresponding to an angular coverage of $0\farcs1 \times 0\farcs1$, corresponding to 69 pc$\,\times$ 69 pc at the galaxy. From now on we will refer to these spatial sampling pixels in the datacube as ``spaxels''.

\section{Results}
\label{res}

\begin{figure*}
\centering
\includegraphics[width=0.8\textwidth]{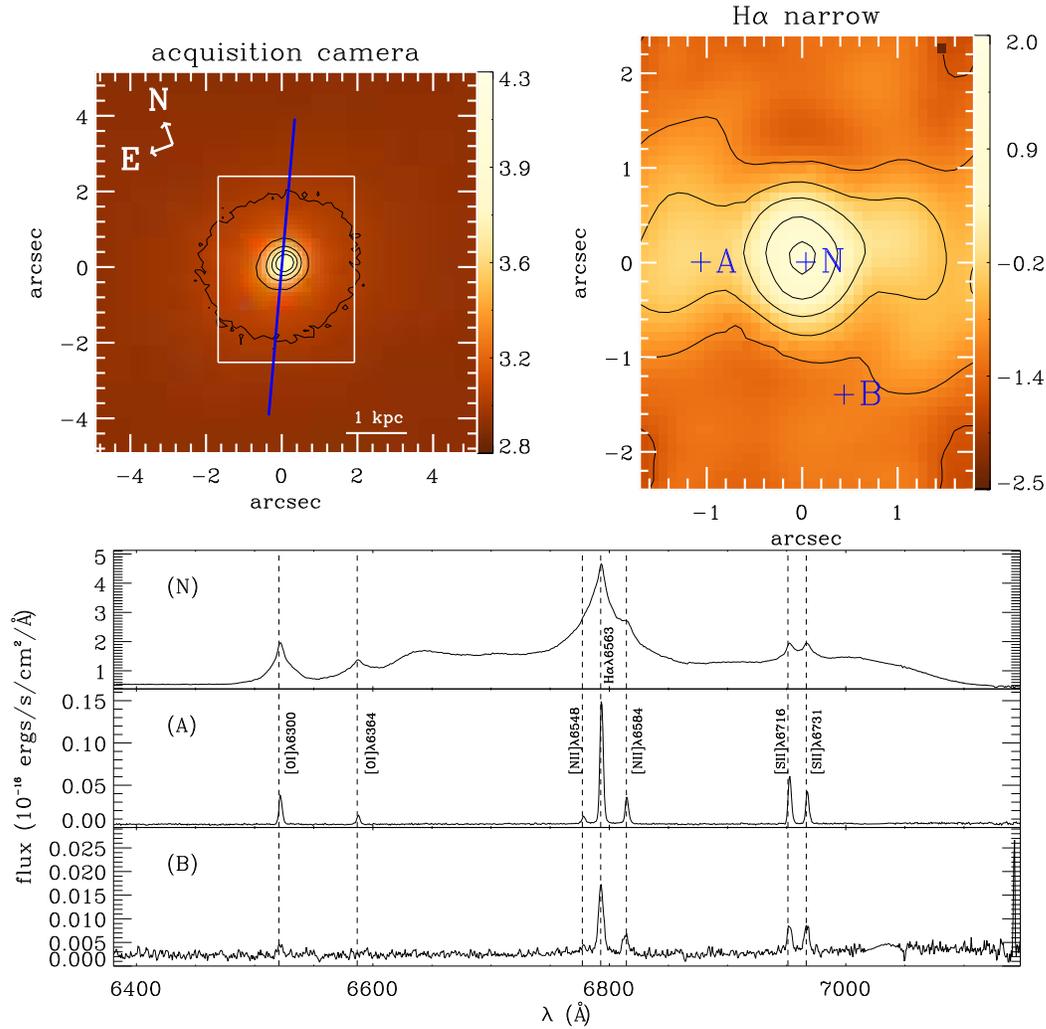}
\caption{Top-left: $\lambda$7800\,\AA\ acquision camera image. The rectangle shows the GMOS IFU FOV. The blue line displays the large-scaled major-axis orientation. The image data are in units of counts and displayed on a logarithmic scale. Top-right: H$\alpha$ flux map from the IFU spectroscopy. Flux units are 10$^{-12}$ erg s$^{-1}$ cm$^{-2}$ arcsec$^{-2}$, and are shown in a logarithmic scale. Bottom: Spectra extracted at the positions N (nucleus), A and B marked at the top-right panel.}
\label{large}
\end{figure*}

Fig. \ref{large} illustrates the IFU data and FOV. In the top-left panel, the acquision image ($\lambda$7060-8500\,\AA, filter i\_G0327) obtained for the GMOS-IFU observation is displayed, covering a region of 10\arcsec\,$\times$\,10\arcsec. The image shows a circular flux distribution with no prominent structures. The blue line shows the major-axis position angle (PA\,$\approx -25^{\circ}$), measured by fitting isophotes in galaxy images obtained from 2MASS and the UK 48-inch Schmidt telescope (central wavelengths of $1.71\,\mu$m and 4680\,\AA; and image sizes of $1.3 \arcmin \times 1.3\, \arcmin\,$ and $3.0 \arcmin \times 3.0\, \arcmin$, respectively). In the top-right panel we present the H$\alpha$ flux map obtained from the IFU spectroscopy. H$\alpha$ emission is observed over the whole field-of-view, with enhanced emission observed along an elongated linear feature (ELF) crossing the nucleus and running from the north-east to the south-west, along the PA $\approx 70^{\circ}$. The ELF apparently extends beyond the IFU FOV, thus more than $1\farcs5$ ($\approx$ 1 kpc) to each side of the nucleus. In the bottom panels we show IFU spectra extracted at the nucleus (position N), $\approx\,1\arcsec$ north-east (position A) and $\approx\,1\farcs4$ south of the nucleus (position B), with positions shown in the top-right panel. The extraction aperture is $0\farcs1 \times 0\farcs1$. The position of the nucleus is defined as the peak of the continuum emission, as measured from an image integrated between 5600\,\AA \, and 6450\,\AA \, (top-left panel of Fig. \ref{flux}). The nuclear spectrum shows the broad double-peaked H$\alpha$ profile, as well as the narrow lines [O\,{\sc i}]$\lambda\lambda$6300,64, [N\,{\sc ii}]$\lambda\lambda$6548,84, H$\alpha\lambda$6563 and [S\,{\sc ii}]$\lambda\lambda$6717,31. The profiles of the narrow lines show some broadening at their bases, which makes it difficult to separate them from the broad double-peaked H$\alpha$ profile at the nucleus. All the narrow lines are also present in the extranuclear spectra, extending up to the edges of the FOV, except for the [O\,{\sc i}] emission lines that are very weak close to the edge of the IFU FOV.

\subsection{Fitting of the emission lines}
\label{fitting}

In order to obtain the integrated emission line flux distributions and gas kinematics, we have fitted the emission-line profiles with Gaussians, using an {\sc IDL}\footnote{Interactive Data Language, http://ittvis.com/idl} routine adapted from the program {\sc PROFIT} \citep{riffel10}. While over most of the FOV a good fit to the narrow emission lines was obtained with a single Gaussian, around the nucleus most of the narrow line profiles presented a broad base, and two Gaussians were necessary. Henceforth, we refer to the component used to fit the broad base of the narrow line profile that is present near the nucleus as the ``broader component''. In addition, all emission lines ``sit on top'' of the broad double-peaked H$\alpha$ profile in the central spaxels. Here the broad line was treated as a local continuum which was fitted using a low order polynomial. Due to the large number of free paramenters in the fit, the following physically motivated constraints were imposed:

\begin{enumerate}
\item different lines from the same ionic species have the same kinematic parameters. For example, the [S\,{\sc ii}]$\lambda\lambda$6717,31 emission lines have the same centroid velocity and velocity dispersion;
\item the [N\,{\sc ii}]$\lambda\lambda$6548,84 emission lines have the same centroid velocity and velocity dispersion as H$\alpha$;
\item the [N\,{\sc ii}]$\lambda$6548 flux was fixed as 1/3 of the [N\,{\sc ii}]$\lambda$6584 flux, in accordance with nebular physics \citep{ostfer06}. This was also done for the [O\,{\sc i}]$\lambda\lambda$6300,64 emission lines.
\end{enumerate}

For most of the narrow emission lines, the two components fit (the narrow plus the broader component) as described above gave the best results for a region around the nucleus with a radius of the order of that of the PSF ($\approx$0\farcs3). An exception is [N\,{\sc ii}], whose lines are particularly faint in Pictor\,A, and whose broader component, if present at all, may be very weak or heavily blended with that of H$\alpha$. For this reason, only the narrow component was used for the [N\,{\sc ii}] lines. We concluded that the broader component, present in the H$\alpha$, [O\,{\sc i}] and [S\,{\sc ii}] emission lines is unresolved in our observations. Outside this region, a single narrow component has been used to fit the lines.

\begin{figure*}
\centering
\includegraphics[width=\textwidth]{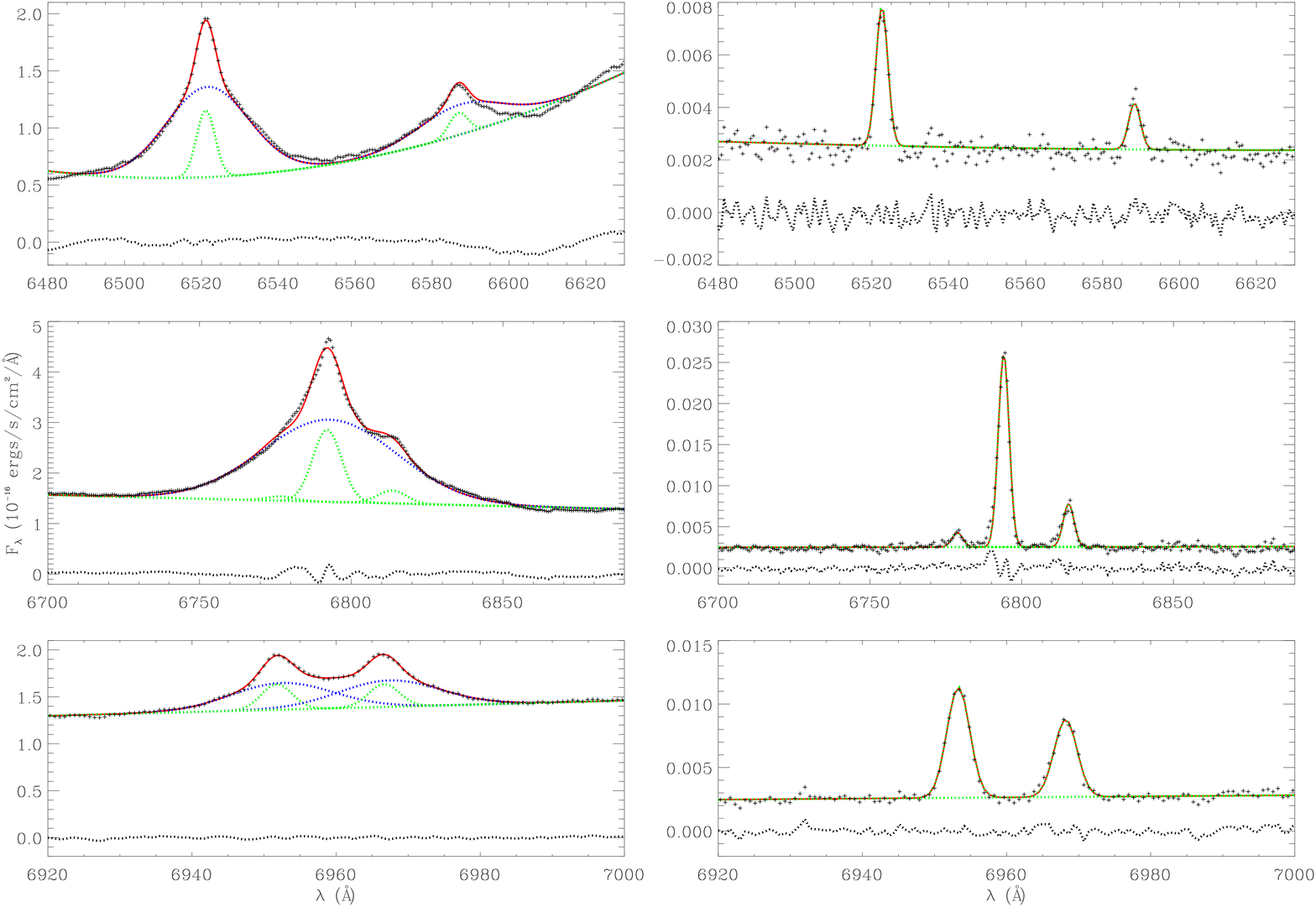}
\caption{Fitting of the nuclear spectrum (left panels) and an example of one extranuclear spectrum taken from a spaxel $\approx 1\farcs2$ from the nucleus (right panels) for emission lines of [O\,{\sc i}] (top panels), H$\alpha$ and [N\,{\sc ii}] (middle panels), and [S\,{\sc ii}] (bottom panels). Each panel displays the narrow (green dotted lines) and broader (blue dotted lines, only in left panels) components, along with the spectrum data points (black crosses), the resulting fit (red line) and the residual spectrum (black dotted lines). Flux units (y axis) are 10$^{-16}$ erg s$^{-1}$ cm$^{-2}$ \AA$^{-1}$.}
\label{fit}
\end{figure*}

Examples of the fits applied to the narrow emission lines of the nuclear and an extranuclear region $\approx 1\farcs2$ west from the nucleus can be seen in Fig. \ref{fit}. The broader component of H$\alpha$ is stronger than in other emission lines, and is certainly blended with a small contribution from the corresponding component of the [N\,{\sc ii}] lines, which may result in a small overestimation of the width and amplitude of the broader H$\alpha$ component.

We used Monte Carlo simulations to estimate the uncertainties on the parameters recovered from the line fits. For each spaxel, we constructed 100 realizations of the spectrum by adding Gaussian noise with amplitude comparable to the noise measured in the original spectrum.

For each realization, Gaussian fits were performed for each emission line. The gaseous mean emission-line integrated flux, centroid velocity and velocity dispersion were then obtained for each emission line, while their respective uncertainties were calculated from the root mean square deviations. The velocity dispersion ($\sigma$) values were corrected for the instrumental broadening.

In order to assess the quality and reliability of the fits, we calculated values of the equivalent width ($EW$), computed for each line in each spaxel, using the mean fit parameters. Inspecting the results for the whole FOV, we reached the conclusion that we could only trust the parameters derived for lines with $EW \ge 4.5$\,\AA. Regions with smaller values of $EW$ for each line were thus masked out from the corresponding maps.

The uncertainties due to the emission-line fitting estimated using the Monte Carlo simulations are small throughout the FOV. Expressing the flux uncertainty, $\epsilon_F$, as a fraction of the integrated flux, $F$, we find that the H$\alpha$, [N\,{\sc ii}] and [S\,{\sc ii}] flux maps show typical values $\epsilon_F / F \approx 0.05\,$ in the nucleus and reaching up to $\epsilon_F / F \approx 0.1-0.15\,$ closer to the border of the FOV. The [O\,{\sc i}] emission lines present higher uncertainty values with $\epsilon_F / F \approx 0.1\,$ in the nucleus and $\epsilon_F / F \approx 0.3\,$ closer to the border of the mask constructed using the equivalent width threshold. The uncertainties of the centroid velocity and velocity dispersion are similar for all emission-lines and present typical values of $\epsilon_v \approx \epsilon_\sigma \approx 10\,$\kms in the nucleus and $\epsilon_v \approx \epsilon_\sigma \approx 20\,$\kms closer to the border of the masked out regions.

\subsection{Emission-line flux distributions} 

\begin{figure*}
\centering
\includegraphics[width=1.0\textwidth]{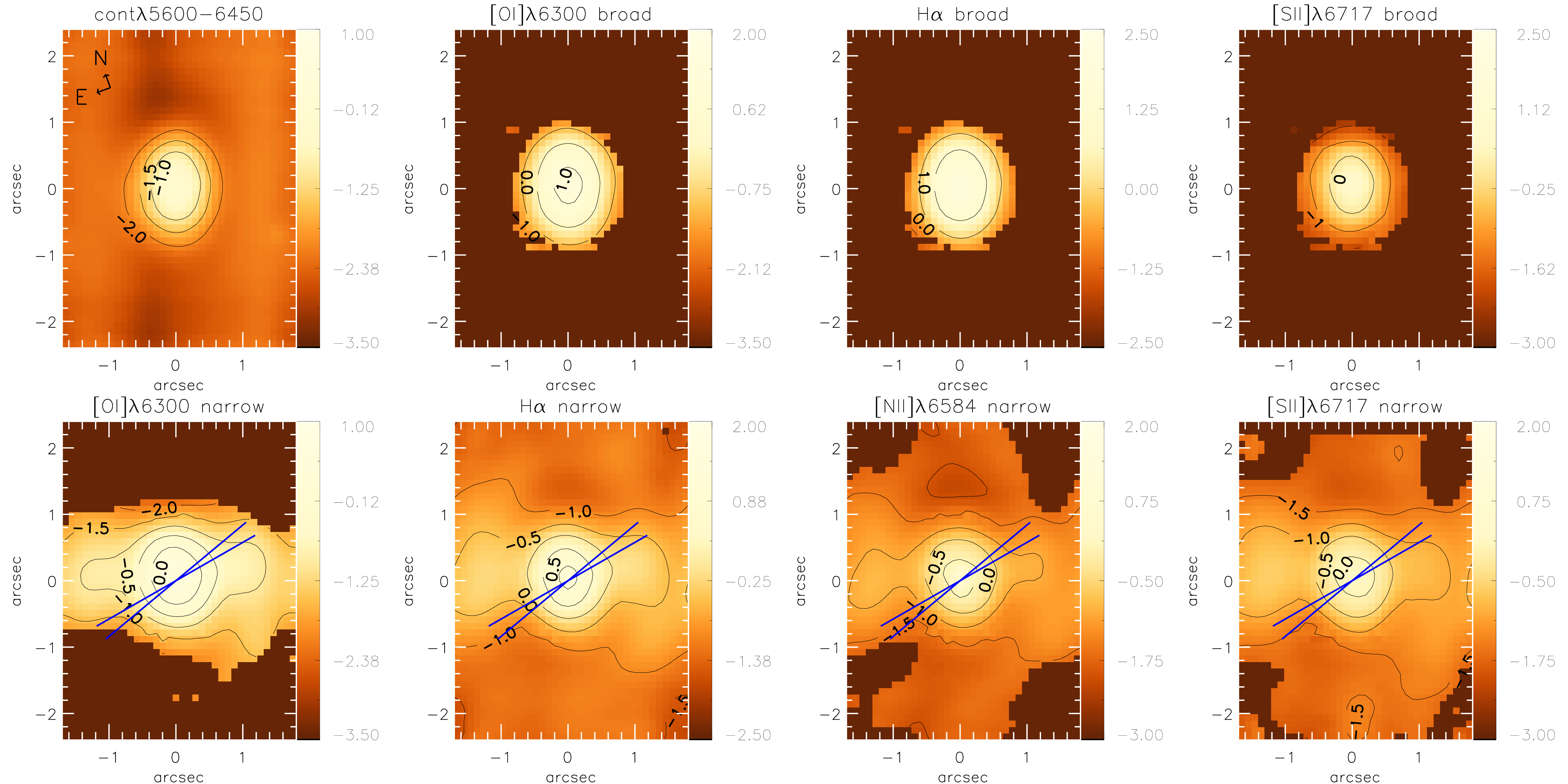}
\caption{Top-left panel: flux map in the continnum. Remaining top panels: flux maps of the broader emission-line components. Bottom panels: flux maps of the narrow emission-line components. The blue lines illustrate the PA range of the large scaled ($\sim 100\,$kpc) radio jet, using values from the literature. Flux units are 10$^{-12}$ erg s$^{-1}$ cm$^{-2}$ arcsec$^{-2}$, and are shown in a logarithmic scale. Flux contours are also plotted and labelled using the same logarithmic scale as the color bar.}
\label{flux}
\end{figure*}

Flux maps were obtained for the emission lines identified in the spectra of Fig. \ref{large}, by integrating the flux in the fitted Gaussians and subtracting the local continuum. They are displayed in Fig. \ref{flux}. {In order to highlight the weaker structures, we use a flux logarithmic scale for the maps presented.} The top-left panel shows a continuum map integrated between $\lambda$5600 and $\lambda$6450, showing almost no structure except the nucleus. Two flux dips are noticeable north and south of the nucleus, which are artifacts of the deconvolution procedure. However, the difference in flux between the dips and the nearby regions corresponds to less than 1\% of the peak of the continuum. These features are only noticeable due to the use of a logarithmic scale for these images. We estimated that the flux difference corresponding to these deconvolution artifacts corresponds to less than 0.2\% of the peak emission in the H$\alpha$ flux maps. The other top panels show the flux distribution of the broader component, centered at the nucleus, that we have concluded is unresolved. The bottom panels show the flux distributions in the narrow component. These display emission in the ELF, crossing the nucleus and running from the north-east to the south-west in PA\,$\approx 70^{\circ}$. This structure appears to extend beyond the limits of our FOV (at $\approx$1.25 kpc) along this PA. Blue solid lines delimit the range of PA's of the radio jet as reported in the literature (PA $= -75 \pm 5^\circ$).

\subsection{Emission-line ratio maps} 
\label{ratios}

\begin{figure*}
\centering
\includegraphics[width=0.8\textwidth]{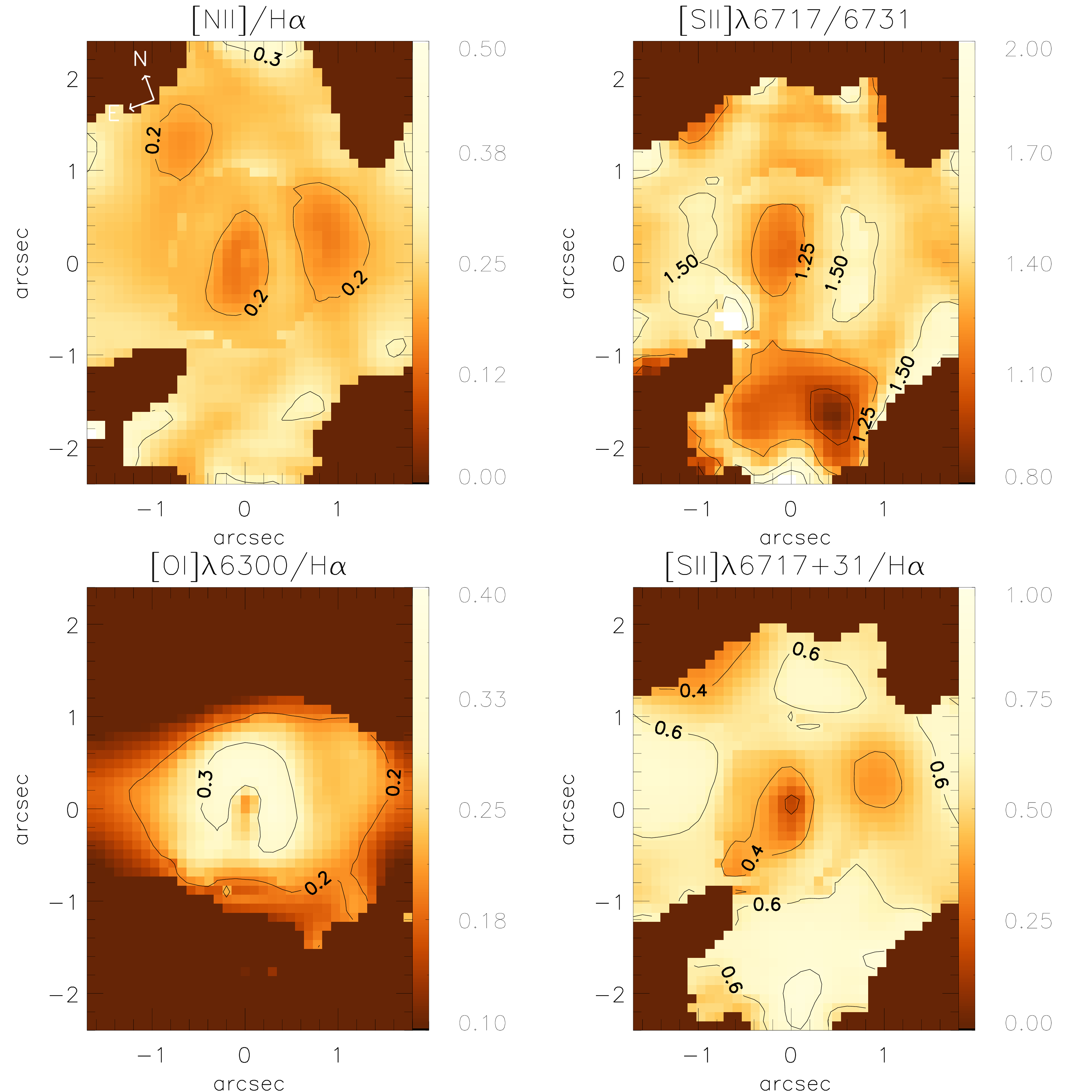}
\caption{Line ratio maps, using only narrow components, for the identified emission lines. Contours represent the flux ratio values for the regions in the maps.}
\label{fluxreason}
\end{figure*}

Fig. \ref{fluxreason} displays line ratio maps obtained from the emission-line narrow component flux distributions. The top-left panel shows the [N\,{\sc ii}]6584/H$\alpha$ ratio map, which indicates remarkably low values around $0.3$ essentially everywhere, with slightly smaller ratios ($< 0.2$) at the nucleus and in two compact regions west and north of the nucleus. The [S\,{\sc ii}]6717/6731 line ratio is displayed in the top-right panel, with values of $\approx 1.3$ characteristic of most of the emission line region. Higher values, of $\approx 1.5$ (at or near the low density limit), are seen along the ELF, while smaller values, of $\approx 1$, are seen south of it. The bottom left panel displays the [O\,{\sc i}]6300/H$\alpha$ ratio, which is highest ($\ge 0.3$) within $\approx$\,0$\farcs$3 of the nucleus and decreases outwards along the ELF. The bottom right panel shows the [S\,{\sc ii}]6717+31/H$\alpha$ ratio, for which the smallest values are found in the nucleus ($\approx 0.25$); it then increases with distance from the nucleus, reaching its highest value ($\approx 0\farcs8$) south of the nucleus. A region of low values ($\approx 0.3$) is also noticeable $\approx 1\arcsec$ west of the nucleus.

Typical uncertainties are of $\approx 10\%$ of the measured value in the regions close to the nucleus, reaching higher values in the unmasked regions closer to the border of the FOV ($\approx 20\%$). The highest unmasked uncertainty values are observed in the regions more distant from the nucleus in the [O\,{\sc i}]6300/H$\alpha$ ratio map, but are smaller than $30\%$ of the ratio value. These uncertainty values were obtained by propagating the flux uncertainties measured in each emission line, as explained in Sec. \ref{fitting}.

\subsection{Gas kinematics}

\subsubsection{Channel maps}

\begin{figure*}
\centering
\includegraphics[width=\textwidth]{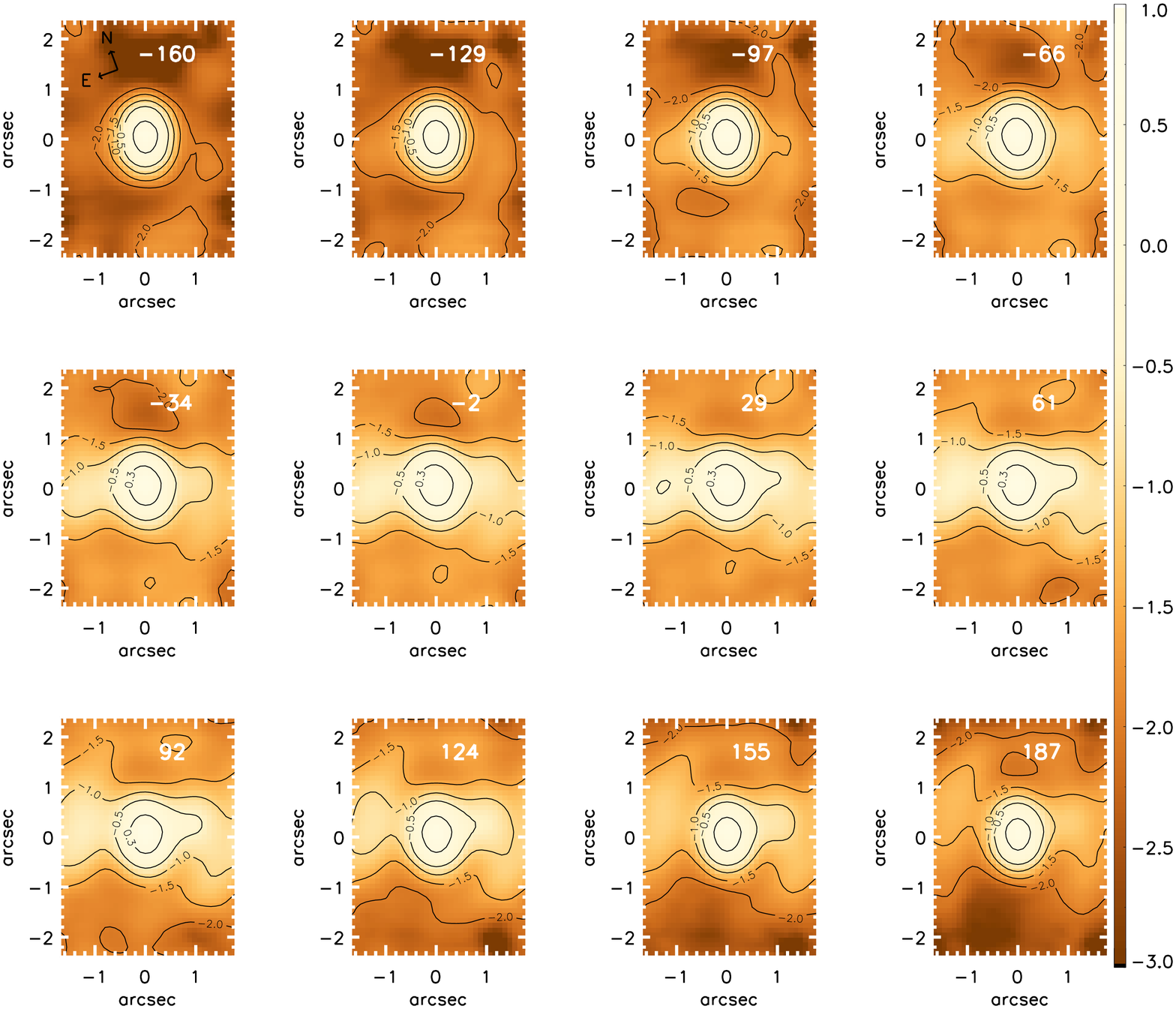}
\caption{Channel maps along the H$\alpha$ emission-line profile, in order of increasing velocities shown in the top of each panel in units of \kms. Flux units are 10$^{-12}$ erg s$^{-1}$ cm$^{-2}$ arcsec$^{-2}$, and are shown in a logarithmic scale. Flux contours are also plotted and labelled using the same logarithmic scale as the color bar.}
\label{cmha}
\end{figure*}

We have mapped the gas kinematics using channel maps. In Fig. \ref{cmha} we display a sequence of these maps extracted within velocity bins of $32$\,\kms (corresponding to one spectral pixel) along the H$\alpha$ emission-line profile. The channel maps reveal blueshifted emission mainly to the S-SE of the nucleus and redshifted emission mainly to the N-NW, as observed also in the centroid velocity maps (Sec. \ref{vel_sig}). However, the most prominent structure is the ELF, dominating the channel maps in the range $v \approx -130$ to $v \approx 180$\, \kms.

\subsubsection{Centroid velocities and velocity dispersions}
\label{vel_sig}

\begin{figure*}
\centering
\includegraphics[width=\textwidth]{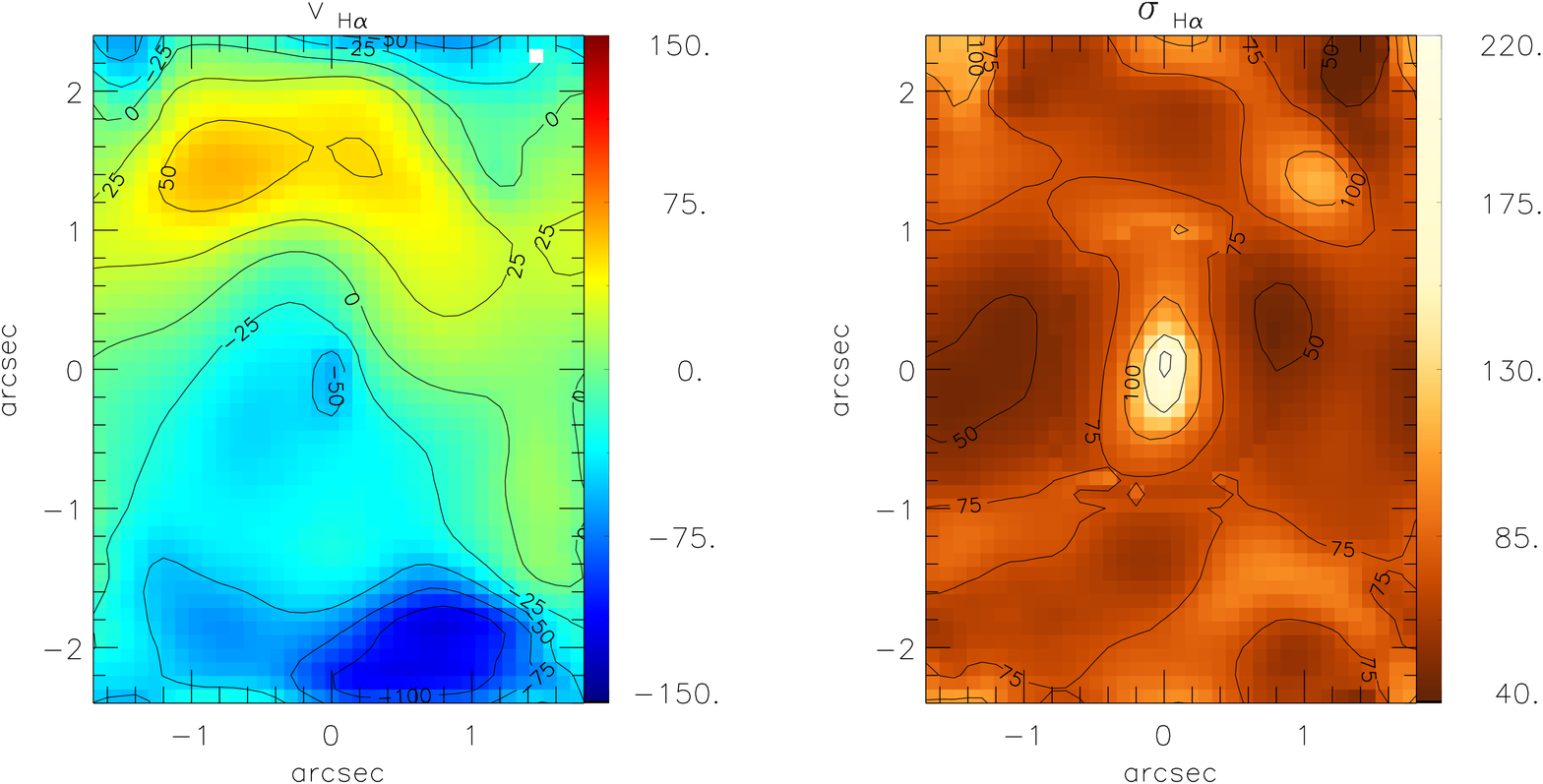}
\caption{Centroid velocity (left panel) and velocity dispersion (right panel) map of the narrow component of the H$\alpha$ emission line. Units are \kms.}
\label{sig}
\end{figure*}

Fig. \ref{sig} shows the centroid velocity map (left panel) and the velocity dispersion map (right panel) for the narrow component of the H$\alpha$ emission line. The heliocentric systemic velocity subtracted from the gaseous centroid velocities is $v_{sys} = 10,526$\,\kms. This value was obtained from the fit of a rotation model to the centroid velocity map of H$\alpha$, as explained in Sec. \ref{kin_dis}, and is in agreement with the systemic velocity from \citet{lauberts89} ($v_{sys} = 10,510 \pm 120$\,\kms), for example. The measured H$\alpha$ peak velocity (integrated through the whole FOV), $v_{\textrm{H}\alpha} = 10,522\pm16\,$\kms, also matches the systemic velocity. The other emission lines present similar kinematical distributions, thus are not shown here.

The centroid velocity map presents blueshifts to the south, with negative velocity values still observed towards the east of the nucleus but decreasing there. Redshifts are found to the north, with positive velocity values still observed towards the west of the nucleus but also decreasing there. A region beyond $2\arcsec$ north-northwest (N-NW) of the nucleus, at the upper border of the map, displays blueshifted velocities beyond the region showing redshifts. We have inspected the fits in this region in order to verify if these blueshifts could be a result of low S/N ratio data, but the H$\alpha$ line is still very strong in this region, with $EW \ge 4.5$\,\AA. 

The mean centroid velocity and velocity dispersion for the broader components of the narrow lines are listed in Table \ref{mean_vel}, where the centroid velocity value is relative to the systemic velocity.

Although the velocity dispersion distributions are very similar between the emission lines, H$\alpha$ exhibits higher values than the other lines. The highest velocity dispersions ($\sigma \approx 200$\,\kms for H$\alpha$; $\approx 100$\,\kms for the other emission lines) of the narrow components are observed at the nucleus. High H$\alpha$ velocity dispersion values ($\approx 100\,$\kms) are also observed along the N-NW border of the map, where blueshifts are found in the centroid velocity maps. Lower values ($\sigma < 50$\,\kms) are observed elsewhere, mainly along the ELF. The broader components show different $\sigma$ values for each emission line, as can be seen in Table \ref{mean_vel}. The velocity dispersion measured in the broader component of H$\alpha\,$ is also larger than those in the other emission lines. As we could not fit this broader component to the [N\,{\sc ii}] emission lines, some contribution from these lines may have been included in H$\alpha$, as discussed above. Assuming, for example, that the broader component in the [NII] lines could be as strong relative to the narrow components as in the [SII] lines, the broader H$\alpha$ would be overestimated by 7.8 $\pm$ 4.5\% in flux. Also, one must keep in mind that the double-peaked broad H$\alpha$ emission may contribute somewhat to the component we identify as the ``broader'' narrow line component.

Also, one must keep in mind that the broad H$\alpha$ double-peaked component may contibute somewhat to the component we label as broader narrow line component. 

\begin{table}
   \centering   
   \caption{\it Nuclear broader components of the narrow lines.}
   \begin{tabular}{|c|c|c|c|c|c|c|c|c|c|} 
      \hline \hline
      Emission line & $\bar{v}\,$ (\kms) & $\bar{\sigma}\,$ (\kms)\\
      $\textrm{[O\,{\sc i}]}\lambda$6300 & $44.0 \pm 3.9$ & $488.0 \pm 3.9$ \\
      H$\alpha$ & $10.6 \pm 2.3$ & $833.2 \pm 7.6$ \\
      $\textrm{[S\,{\sc ii}]}\lambda$6717 & $-17.2 \pm 2.5$ & $237.3 \pm 3.9$ \\
      \hline
   \end{tabular}
   \label{mean_vel}
\end{table}

\section{Discussion and interpretation}
\label{disc}

\subsection{Flux distributions}

The narrow components of all the emission lines shown in Fig. \ref{flux} present similar flux distributions dominated by the ELF that extends from the nucleus to the W-SW and E-NE borders of the FOV ($\approx 1\farcs7 \approx 1.2\,$kpc). This structure is oriented along a direction close to but not aligned with that of the radio jet axis reported in the literature. The emission seems to extend beyond the limits of our FOV at lower flux levels. We note that the surface brightness from the ELF is $\sim 30\,$ times lower than the nuclear emission, when considering just the narrow component. The broader line components originate from a spatially unresolved region characterized by a high velocity dispersion probably due to complex kinematics that could include outflows from the AGN. Fainter emission extends south-east and north-west of the ELF, covering the entire FOV in H$\alpha$.

\subsubsection{Gas excitation} 
\label{gasexc}

\begin{figure*}
\centering
\includegraphics[width=\textwidth]{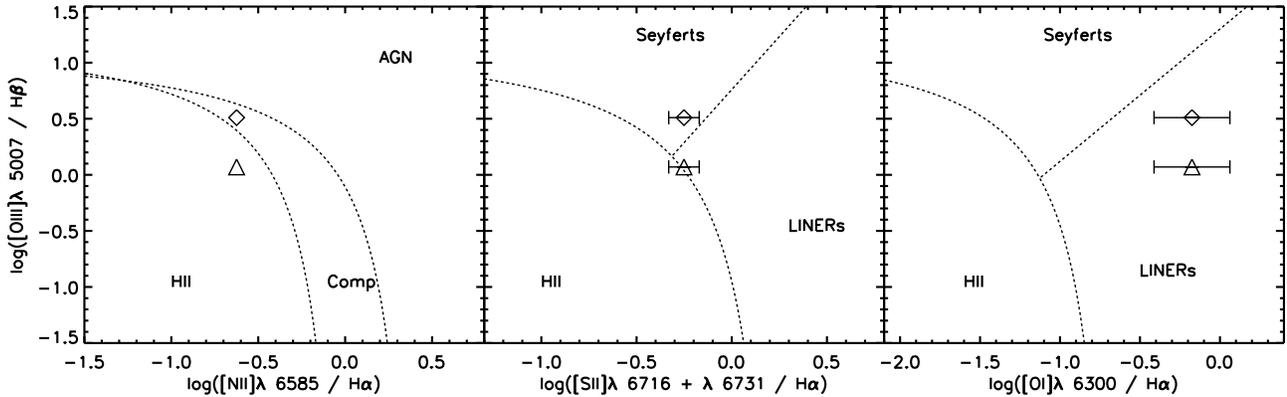}
\caption{BPT diagrams for the mean flux ratios of Pictor A over the whole FOV. Values for the ratio [O\,{\sc iii}]/H$\beta$ are from \citet[triangles]{filippenko85} and \citet[diamonds]{simkin99}. Horizontal error bars represent the standard deviation of the ratio values across the FOV, and are not shown if smaller than the symbol sizes. The dashed boundary lines are taken from \citet{kewley06}. ``Comp'' stands for composite galaxies.}
\label{bpt}
\end{figure*}

Fig. \ref{bpt} displays the line ratios derived from the narrow components of the emission lines plotted in the Baldwin-Phillips-Terlevich (BPT) diagrams \citep{baldwin81}. Only the mean value for the whole FOV is plotted for each line ratio, with its standard deviation being represented by the error bars. As our spectral coverage does not include the H$\beta$ and [O\,{\sc iii}]$\lambda$5007 emission lines, we used the ratio values reported by \citet[diamonds]{simkin99} and \citet[triangles]{filippenko85}. \citet{simkin99} report values of $log$ ([O\,{\sc iii}]/H$\beta$) ranging from $0.46$ to $0.56$, measured along two slits oriented at PAs $102^\circ$ and $128^\circ$, which begin at the nucleus and end at $2.2 \arcsec$ W of the nucleus. \citet{filippenko85}, using an integrated spectrum (aperture $2 \arcsec \times 4 \arcsec$), reports a much smaller ratio of $\approx 0.07$. The [N\,{\sc ii}]6584/H$\alpha$ ratio values (left panel) are quite small and are typical of H{\sc ii} regions. As shown in Fig. \ref{bpt}, the [N\,{\sc ii}]6584/H$\alpha$ ratio deviation is small ($\sigma = 0.04$, smaller than symbol size).

The second BPT diagram (central panel), involving [S\,{\sc ii}]/H$\alpha$, shows line ratio values in the border between H{\sc ii} regions and AGNs when the [O\,{\sc iii}]/H$\beta$ is taken from \citet{filippenko85}, but typical of Seyferts for the \citet{simkin99} value of [O\,{\sc iii}]/H$\beta$. On the other hand, [O\,{\sc i}]6300/H$\alpha$ values are typical of LINERs, independently of the two previous works, and this may indicate the presence of shocks or dilute AGN radiation. The weakness of [N\,{\sc ii}] and [S\,{\sc ii}] emission lines when compared to H$\alpha$ is noticeable when inspecting the spectra (see for example Fig. \ref{large}). This was also noted by \citet{simkin99}, who report line ratios similar to the ones we derive. These authors mention that gas clouds in sources associated with radio jets are reported to present low [N\,{\sc ii}]6584/H$\alpha$ ratios \citep{clark97,villar98}. However, the case of Pictor A seems extreme, since it displays even lower [N\,{\sc ii}]6584/H$\alpha$ ratios than the objects discussed in these works. Even though the [N\,{\sc ii}]6584/H$\alpha$ ratio is typical of H{\sc ii} regions, we do not find any other indication of the presence of a young stellar population either in our spectra or in the previously published data \citep{filippenko85,halera94}.

We have estimated the Oxygen abundance (O/H) for the emitting gas in Pictor A using the relations of \citet{storchi98}. Adopting the [O\,{\sc iii}]/H$\beta$ values from \citet{simkin99} and \citet{filippenko85} we obtain an average $12 + \textrm{log}\,(O/H) = 8.39 \pm 0.02$ over the whole FOV. This value is quite low, considering that the narrow-line region of active galaxies usually have metallicities of 2-3 times the solar value \citep{storchi89,tremonti04,groves06}. A similar low abundance is implied by the relation between [N\,{\sc ii}]/H$\alpha$ and metallicity presented by \citet{kewley13}. An oxygen abundance of $12 + \textrm{log}\,(O/H) = 8.39$ corresponds to $\approx -0.9 <$ log ([N\,{\sc ii}]/H$\alpha$) $< -1.5$, according to the models. One possibility is that the sub-solar metallicity observed in the circumnuclear region of Pictor A may be a consequence of the merger event related to the tidal tail reported by \citet{gentry15}, in which a low-metallicity gas rich galaxy was accreted, leading to an accumulation of low-metallicity gas in the region covered by our FOV. It has been argued that mergers are more common in radio galaxies than in their quiescent counterparts \citep{ramos11}, and we may be observing the outcome of such an event. The lower [N\,{\sc ii}]/H$\alpha$ ratio values (left upper panel of Fig. \ref{fluxreason}), may be tracing regions where this low-metallicity gas is most concentrated.

\subsubsection{Electron density}

\begin{figure*}
\centering
\includegraphics[width=0.8\textwidth]{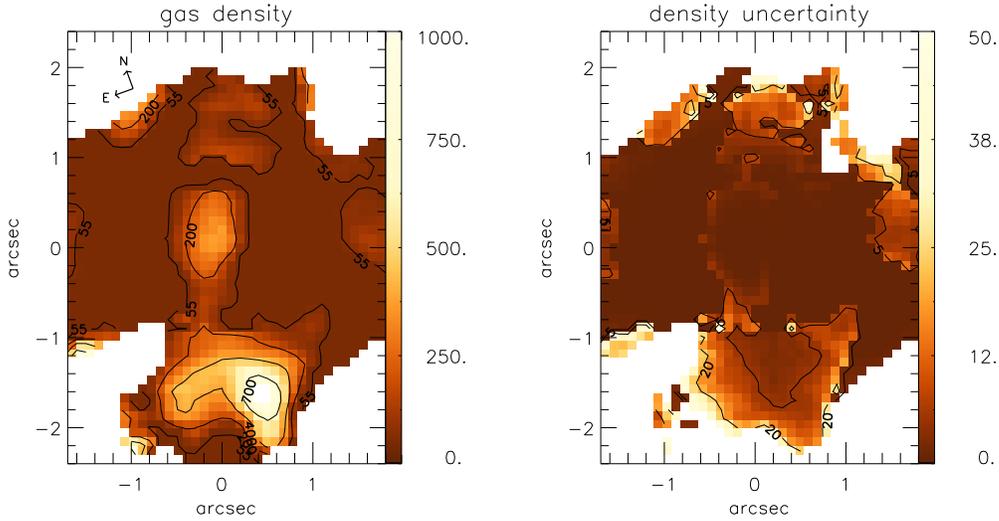}
\caption{Left: electron density map, derived from the [S\,{\sc ii}] ratio, in units of cm$^{-3}$. Right: percent uncertainty in the density values.}
\label{dens}
\end{figure*}

The [S\,{\sc ii}] ratio map (top-right panel in Fig. \ref{fluxreason}) was used to derive the electron density map \citep{ostfer06}, shown in the left panel of Fig. \ref{dens}. In the right panel we display the uncertainty map of the electron density, propagated from the flux uncertainties measured using the Monte Carlo iterations, in percent scale. The temperature used to derive the electron density is 10,000\,K. The average gas density calculated over the entire IFU FOV is $n_e = 126^{+49}_{-23}\,$cm$^{-3}$. There are significant ($> 3\sigma$) variations in the [S\,{\sc ii}]6731/6717 ratio over the FOV, implying corresponding variations in density.

The highest density values ($> 700\,$cm$^{-3}$) are found in a region $\approx 1\farcs8$ ($1.2\,$kpc) south of the nucleus. The lowest electron densities ($< 100\,$ cm$^{-3}$) are observed to the north-east and south-west of the nucleus along the ELF. High [S\,{\sc ii}] line ratio values are observed in this region, and in some pixels of the map the ratios are higher than that corresponding to the low-density limit of the [S\,{\sc ii}] line ratio. In these pixels we fixed the density to a value of $50\,$ cm$^{-3}$, with uncertainties derived directly from the line ratio. At the nucleus, the density is somewhat higher ($\approx 300\,$cm$^{-3}$). The uncertainties in the density values are usually below $30\%$ of the estimated electron density over the whole FOV. We have also calculated the electron density related to the broader emission component. We fitted the [S\,{\sc ii}] emission lines in an integrated spectrum extracted from the IFU data, masking out regions where the broader component is not present. We obtain a density for the broader component of $n_e = 871 \pm 1\,$cm$^{-3}$.

\subsection{Gas kinematics}
\label{kin_dis}

Fig. \ref{sig} shows that the centroid velocity field suggests some rotation: mostly blueshifts to the south and redshifts to the north. However, the zero velocity curve is quite distorted, which is not expected in pure rotation kinematics. Another component is observed $\approx 2 \arcsec\,$ to the N-NW of the nucleus, with blueshifted velocities further from the redshifted region. In addition, blueshifts observed approaching the nucleus from the east and redshifts from the west appear to correspond to yet another, more subtle third kinematical component, probably related to the distorted zero velocity contour. This component may also be related to the ELF (see Fig. \ref{flux}), as the orientation is approximately the same. 

These features are broadly consistent with the larger scale ($\sim 10 \arcsec \times 10 \arcsec$) velocity map constructed by \citet{simkin99}. Our FOV covers only the central region of their reconstructed velocity map. However, the redshifted and blueshifted velocities observed to the north-east and south-west of the nucleus, respectively, and the redshifted velocity region extending up to $\approx 2.5 \arcsec$ west of the nucleus (see their Fig. 5), can be readily identified with the features outlined above that are seen in our maps. The red- and blueshifted regions that appear at distances $> 2 \arcsec$ east and west of the nucleus in the \citet{simkin99} map are outside of our FOV.

The channel maps shown in Fig. \ref{cmha} present both negative and positive velocity values along the ELF, suggesting that it is approximately in the plane of the sky. This would produce small velocity amplitudes in this region, which is what we observe in the centroid velocity maps, in agreement with this scenario. Perpendicular to the ELF, the brightest extranuclear emission in blueshift is found to the south whereas the brightest extranuclear emission in redshift is found to the north of the nucleus. This north-south orientation is not far from that of the large scale galaxy major axis, shown in Fig. \ref{large}, thus the kinematics seems to be consistent with rotation in the galaxy. One may speculate that the merger scenario mentioned in Sec. \ref{gasexc} would result in this kinematical component, that drives gas into the ELF. 

We have tried to fit a disk rotation model to the H$\alpha$ centroid velocity field. This model \citep{bertola91} has been used by our group to model the velocity field in other galaxies that present rotation, such as NGC 2110 \citep{schnorr14} and NGC 1386 \citep{lena15}. The velocity residuals resulting from the fit were comparable to the velocity field itself, with ``non-rotating" components showing velocities as high as those that could be attributed to rotation. The residuals are mainly observed in redshifted and blueshifted velocities along the west and east regions of the nucleus, respectively, as well as blueshifts to the N-NW at the borders of the FOV. We concluded that the rotation model is not a good representation of our observations, probably due to the complex kinematics, thus this model is not discussed further. However, as noted in Sec. \ref{vel_sig}, we adopted the systemic velocity derived from the best fit rotation model, since it did not change considerably between fits, and is also in agreement with the literature.

Velocity dispersions are low outside the nucleus, being typically smaller than $100\,$\kms. The most interesting feature observed in the right panel of Fig. \ref{sig} is the even smaller velocity dispersions along the ELF. This seems to be inconsistent with the channel maps that show the ELF in several velocity bins. However, inspecting the line fitting in this region, we noticed wings in the residuals, which are responsible for the presence of the ELF in several velocity bins shown in the channel maps (Fig. \ref{cmha}). The residuals have flux intensities lower than $5\%$ of the line peak flux, and are only noticeable due to the use of a logarithmic scale. This low velocity dispersion within the ELF is strong evidence against the scenario in which the ELF originates from a jet-cloud interaction. Such interactions would be expected to produce higher velocity dispersions of typically $200\,$\kms or above \citep[e.g.][]{barbosa09,storchi10}. We discuss this scenario further in the next sections. 

\subsubsection{Principal Component Analysis}

\begin{figure*}
\centering
\includegraphics[width=\textwidth]{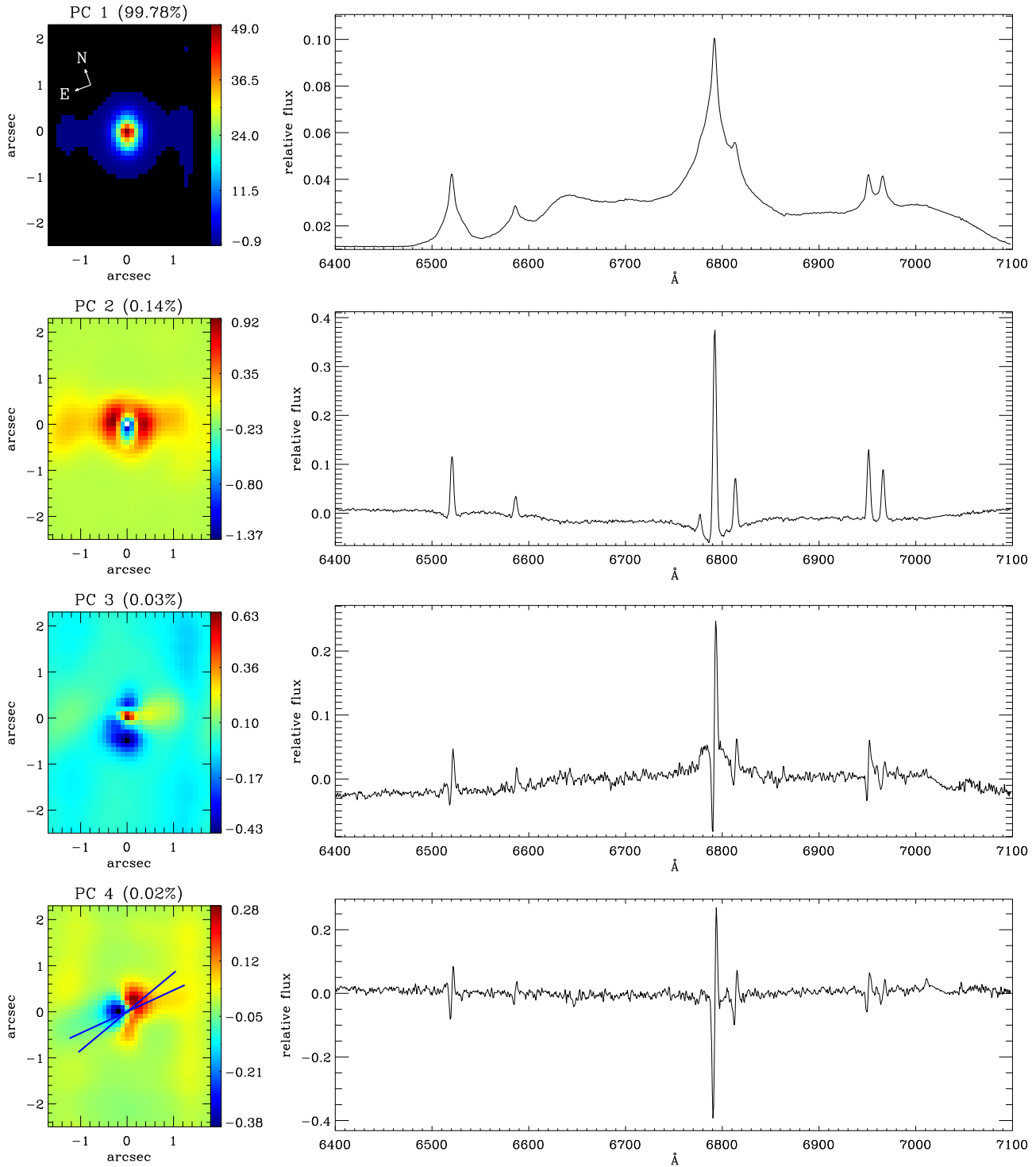}
\caption{PCA analysis. From top to bottom: tomograms (left panels) and eigenspectra (right panels) corresponding to PC1, PC2, PC3 and PC4, in descending order of variance.}
\label{pca}
\end{figure*}

In order to try to obtain further insight into the nature of the circumnuclear gas motions in Pictor A, we have applied Principal Component Analysis (PCA) to the IFU data \citep{steiner09}, following a method similar to that used in previous studies by our group \citep{schnorr11,couto13}. The PCA was applied directly to the calibrated datacube after the Richardson-Lucy deconvolution. The method can be regarded as a separation of the information that was originally in a system of correlated coordinates (in this case, the IFU datacube), resulting in a system of uncorrelated coordinates ordered by principal components of decreasing variance. These new coordinates are eigenvectors which are functions of wavelength. They are called ``eigenspectra'', and reveal spatial correlations and anti-correlations associated with the emission-line features. The projection of the data onto these new coordinates produces images called ``tomograms'', which represent the spatial distribution of the eigenspectra.

The results of the PCA are shown in Fig. \ref{pca}. From top to bottom the panels show the first four tomograms (left panels) and corresponding eigenspectra (right panels), in order of decreasing variance.

Component PC1 accounts for almost all the variance in the data ($99.78\%$) and displays an eigen-spectrum dominated by the AGN and the broad double-peaked H$\alpha$ profile, along with the narrow emission-lines. These are represented by positive values in the eigenspectrum. The spatial distribution comprises the unresolved nucleus, origin of the double-peaked broad H$\alpha\,$ line and broader component of the narrow lines, and the ELF, that is the origin of the brightest emission of the narrow component of the narrow lines.

Component PC2 displays an anti-correlation between broad double-peaked and narrow emission-line components. The broad double-peaked component appears with negative values in the eigenspectrum (represented in blue in the tomogram), while the narrow components show positive values (red in the tomogram). In the corresponding tomogram one can observe that the gas responsible for the narrow line emission comes from an extranuclear region, extending along the ELF. The broad component seems to include not only the double-peaked component but also the broader component of the narrow emission lines, and is confined to the nucleus, according to the tomogram, being anti-correlated with the extended narrow-line emission along the ELF.

Component PC3 shows a small positive contribution associated with the broad double-peaked H$\alpha$ line. The narrow emission lines show a negative contribution in blueshift (appearing as an absorption in the eigenspectrum), thus anti-correlated with the positive contribution of the double-peaked component, but also show a positive contribution in redshift (appearing as emission in the eigenspectrum). In the tomogram, the double-peaked component appears confined to the nucleus; the redshifted component of the narrow line appears along the ELF mostly to the south-west, while the blueshifted component appears almost perpendicular to this structure, with most emission to the south-east. It is unclear how the information in PC3 relates to any other characteristic we have found in our previous analysis.

\begin{figure}
\centering
\includegraphics[width=0.45\textwidth]{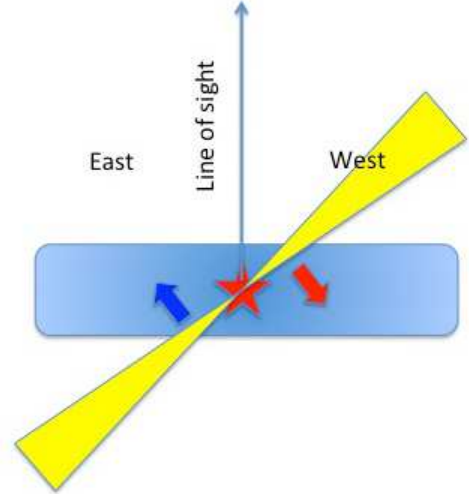}
\caption{A schematic diagram of the possible origin of the PC4 component. In this scenario, lateral expansion of hot gas heated by shocks at the interface between the jet (shaded  yellow) and the ambient medium gives rise to the observed blueshifts and redshifts. As the jet emerges from the ELF (blue) at a relatively steep angle, redshifts occur on the near (west) side and blueshifts on the far (east) side.}
\label{jet_scheme}
\end{figure}

Component PC4 does not show any contribution from the broad double-peaked component, but shows an anti-correlation between blueshifted and redshifted parts of the narrow-line profiles. This is represented by two opposite regions -- relative to the nucleus -- in the tomogram, with an orientation approximately coincident with that of the radio jet. This may indicate an interaction between the gas emitting these features and the radio jet. The redshifted region in PC4, however, is located to the west of the nucleus, and therefore corresponds to the near-side (approaching) jet, as indicated by the fact that both the X-ray jet \citep{wilson01} and the parsec-scale radio jet \citep{tingay00} are on the western side of the nucleus. One might naively expect a jet-gas interaction to generate blueshifts, rather than redshifts, on the approaching side and vice versa. However, Pictor A is an FR II radio source and the two jets have long since ``punched out'' of the galaxy and now terminate at hot-spots \citep{tingay08} in their respective lobes far outside our FOV \citep[the lobes are 4-5$\arcmin$ from the nucleus;][]{perley97}. Therefore, any interaction with the ELF seems likely to involve lateral expansion of gas around the jet, perhaps driven by hot gas heated by shocks at the interface between the jet and the ambient medium. As the jet axis is not aligned with the ELF, and there is a larger column of gas below the jet on the west side, and above it on the east side (particularly, if there is a vertical density gradient in the ELF), lateral expansion of gas surrounding the jet may result in the observed pattern of blueshifts and redshifts in PC4 (Fig. \ref{jet_scheme}).

An alternative possibility is that this component is tracing rotation in a compact disk. The centroid velocity field shows redshifted and blueshifted velocities west and east of the nucleus, respectively, which may be related to the feature observed in PC4. However, considering that the jet probably intersect the ELF (see Sec. \ref{incjet}), we cannot distinguish between these two scenarios, and we consider the origin of this component to be uncertain.

\subsection{Inclination of the radio jet}
\label{incjet}

In order to better understand the possible correlation of the radio jet with the flux distributions and kinematics of the ionized gas, we have attempted to obtain an estimate of the jet orientation angle ($\theta_j$), the angle between the radio jet axis and the line of sight. We used information on apparent jet speed relative to the speed of light ($\beta_{app}$) and jet-to-counterjet surface brightness ratio ($R$) from the parsec-scale data presented in \citet{tingay00}. In this work, several components of the milliarcsecond-scaled jet were identified using VLBI observations. Since component C3 (see their Fig. 2), which is approximately 5 milliarcseconds ($=3.4\,$ parsec) from the parsec-scale core, has the smallest relative error in its measured $\beta_{app}$, we used this jet component to estimate $\theta_j$. Using $R>27$, $\beta_{app}=1.1$, jet structural parameter $p=3$ (for a continuous jet), and the relations listed in the Appendix of \citet{urry95}, we derive an upper limit of $\theta_j \leq 48$ degrees. This result is consistent with the inclination estimated by \citet{tingay00}, who derived the jet inclination in a different manner, using the jet deflection model for a bent jet following \citet{conway93}, finding an upper limit of $\theta_j<51^\circ$. Another estimate by \citet{wilson01}, incorporating projection relations from the radio maps of \citet{perley97} yields $\theta_j=23^\circ$.

The jet inclination we find is an upper limit, but far from being close to the plane of the sky. Even though the radio jet extends to great distances ($\approx 400\,$ kpc overall), the inclination derived is relatively small. The fact that only one side of the X-ray jet is visible also favors a small inclination \citep{wilson01}. However, the ELF observed here seems to be close to the plane of the sky, as discussed in Sec. \ref{kin_dis}, suggesting that it does not have the same orientation as the radio jet. If so, the origin of this feature is probably unrelated to the radio jet, arguing against shocks as a main mechanism of ionization of the gas. This is also consistent with the low velocity dispersion values observed along the ELF. With that said, one must keep in mind that there may still be some interaction of the radio jet with the gas. The outflow related to the radio jet may partially intersect the gas streaming along the ELF at some inclination and result in the feature we are observing in PC4. Another possibility is a change of direction of the jet, when comparing parsec to kiloparsec scales. Such behaviour has been observed in other FRII sources \citep{kharb10}.

\section{Conclusions}
\label{conc}

Using the GMOS integral field unit we observed the inner $2.5\,$kpc $\times\,3.4\,$kpc of the broad-line radio galaxy Pictor A, with a spatial resolution of $\approx 400\,$pc at the galaxy. We modeled the profiles of the emission lines (from [O\,{\sc i}]6300 to [S\,{\sc ii}]6717,31) to produce integrated flux, centroid velocity and velocity dispersion maps. The main conclusions from our measurements and analysis are:

\begin{enumerate}
\item two components were fitted to the narrow emission lines: one narrow ($\sigma \approx 100\,$\kms) and another broader ($\sigma > 300\,$\kms) component, which was only observed at the nucleus and was spatially unresolved by our observations;
\item besides the strong nuclear emission, the most remarkable feature is an elongated linear feature (ELF) crossing the nucleus and running from the south-west to the north-east, along PA $\approx 70^\circ$, up to the borders of the FOV ($\approx$1.2 kpc from the nucleus);
\item the [O\,{\sc i}]6300/H$\alpha$ ratio values are typical of AGNs, in contrast with [N\,{\sc ii}]6584/H$\alpha$, which presents values typical of H {\sc ii} regions; we propose that the low [NII]/H$\alpha$ ratios are due to the low metallicity of the gas ($12 + \textrm{log}\,(O/H) = 8.39 \pm 0.02$);
\item the gas centroid velocity maps display mostly blueshifts to the south and redshifts to the north, but the velocity field is not accurately described by a circular rotation model due to deviations observed close to the nucleus and at the northern border of the FOV;
\item the gas velocity dispersions are high at the nucleus ($\sigma \approx 200\,$\kms) and low ($< 100\,$\kms) elsewhere, including along the ELF, arguing against a jet-cloud interaction as the origin of the ELF even though the radio jet orientation on the sky is quite close to the major axis of the ELF;
\item the ELF seems to be unrelated to the radio jet and displays co-spatial blueshifts and redshifts in the channel maps, suggesting that it is in the plane of the sky;
\item the Principal Component Analysis PC4 component displays a compact bi-polar structure ($< 400\,$ pc from the nucleus), which we interpret as a jet-ELF interaction although rotation in a compact disk cannot be ruled out.
\end{enumerate}

Based on the above conclusions, we propose a scenario for Pictor A in which the ELF is being ionized by the radiation field of the active nucleus. The ELF does not seem to have a stellar counterpart nor is it the product of a jet-cloud interaction. As the gas has a low metallicity, we suggest it may come from a low-metallicity companion galaxy that was captured by Pictor A. The kinematics of the faint emission line gas beyond the ELF suggests that it may be connected to -- perhaps feeding -- the ELF. Feeding of the AGN could be achieved through the ELF, driving the gas captured in the merger to the galaxy nucleus, initiating and sustaining its activity. It seems that the radio jet of Pictor A does not have much influence on the line emitting circumnuclear gas throughout the few inner kpcs analyzed in this work, neither energetically nor kinematically, even though we do trace a possible weak jet-gas interaction through the PCA analysis. We then conclude that within our FOV, we do not observe significant feedback related to the radio jet. However, strong jet-medium interaction on larger scales cannot be ruled out from our observations. Indeed, jet stratification possibly arising due to strong jet-medium interaction has been suggested for the X-ray jet of Pictor A by \citet{hardcastle15}.

\section*{Acknowledgments}

We thank the referee for relevant suggestions which helped to improve this paper. This work is based on observations obtained at the Gemini Observatory, which is operated by the Association of Universities for Research in Astronomy, Inc., under a cooperative agreement with the NSF on behalf of the Gemini partnership: the National Science Foundation (United States), the National Research Council (Canada), CONICYT (Chile), the Australian Research Council (Australia), Minist\'{e}rio da Ci\^{e}ncia, Tecnologia e Inova\c{c}\~{a}o (Brazil) and Ministerio de Ciencia, Tecnolog\'{i}a e Innovaci\'{o}n Productiva (Argentina). This work has been partially supported by the Brazilian institution CNPq. This material is based upon work supported in part by the National Science Foundation under Award No. AST-1108786.

\label{lastpage}


\begin{thebibliography}{2}

\bibitem[\protect\citeauthoryear{Allington-Smith et al.}{2002}]{allington02} Allington-Smith, J. R., Murray, G., Content, R., Dodsworth, G., Davies, R., Miller, B. W., Jorgensen, I., Hook, I., Crampton, D., Murowinski, R., 2002, PASP, 114, 892

\bibitem[\protect\citeauthoryear{Baldwin et al.}{1981}]{baldwin81} Baldwin, J. A., Phillips, M. M., Terlevich, R., 1981, PASP, 93, 5

\bibitem[\protect\citeauthoryear{Barbosa et al.}{2009}]{barbosa09} Barbosa, F. K. B., Storchi-Bergmann, T., Cid Fernandes, R., Winge, C., Schmitt, H., 2009, MNRAS, 396, 2

\bibitem[\protect\citeauthoryear{Baum \& Heckman}{1989}]{bauhec89} Baum, S. A., Heckman, T., 1989, ApJ, 336, 702


\bibitem[\protect\citeauthoryear{Bertola et al.}{1991}]{bertola91} Bertola, F., Bettoni, D., Danziger, J., Sadler, E., Sparke, L., de Zeeuw, T., 1991, ApJ, 373, 369

\bibitem[\protect\citeauthoryear{Best et al.}{2000}]{best00} Best, P. N., R{\"o}ttgering, H. J. A., Longair, M. S., 2000, MNRAS, 311, 23


\bibitem[\protect\citeauthoryear{Clark et al.}{1997}]{clark97} Clark, N. E., Tadhunter, C. N., Morganti, R., Killeen, N. E. B., Fosbury, R. A. E., Hook, R. N., Siebert, J., Shaw, M. A., 1997, MNRAS, 286, 558

\bibitem[\protect\citeauthoryear{Conway \& Murphy}{1993}]{conway93} Conway, J. E., Murphy, D. W., 1993, ApJ, 411, 89

\bibitem[\protect\citeauthoryear{Couto et al.}{2013}]{couto13} Couto, G. S., Storchi-Bergmann, T., Axon, D. J., Robinson, A., Kharb, P., Riffel, R. A., 2013, MNRAS, 435, 2982

\bibitem[\protect\citeauthoryear{Di Matteo et al.}{2005}]{dimatteo05} Di Matteo, T., Springel, V., Hernquist, L., 2005, Nature, 433, 604


\bibitem[\protect\citeauthoryear{Eracleous \& Halpern}{2003}]{erahal03} Eracleous, M., Halpern, J. P., 2003, ApJ, 599, 886

\bibitem[\protect\citeauthoryear{Fanaroff \& Riley}{1974}]{fanril74} Fanaroff, B. L. , Riley, J. M., 1974, AJ, 79, 745

\bibitem[\protect\citeauthoryear{Ferrarese \& Merritt}{2000}]{fermer00} Ferrarese, L., Merritt, D., 2000, ApJ, 539, 9

\bibitem[\protect\citeauthoryear{Feruglio et al.}{2010}]{feruglio10} Feruglio, C., Maiolino, R., Piconcelli, E., Menci, N., Aussel, H., Lamastra, A., Fiore, F., 2010, A\&A, 518, 155

\bibitem[\protect\citeauthoryear{Filippenko}{1985}]{filippenko85} Filippenko, A. V., 1985, ApJ, 289, 475

\bibitem[\protect\citeauthoryear{Fragile et al.}{2004}]{fragile04} Fragile, P. C., Murray, S. D., Anninos, P., van Breugel, W., 2004, ApJ, 604, 74

\bibitem[\protect\citeauthoryear{Gebhardt et al.}{2000}]{gebhardt00} Gebhardt, K., Bender, R., Bower, G., Dressler, A., Faber, S. M., Filippenko, A. V., Green, R., Grillmair, C., Ho, L. C., Kormendy, J., Lauer, T. R., Magorrian, J., Pinkney, J., Richstone, D., Tremaine, S., 2000, ApJ, 539, 13

\bibitem[\protect\citeauthoryear{Gentry et al.}{2015}]{gentry15} Gentry, E. S., Marshall, H. L., Hardcastle, M. J., Perlman, E. S., Birkinshaw, M., Worrall, D. M., Lenc, E., Siemiginowska, A., Urry, C. M., 2015, ApJ, 808, 92

\bibitem[\protect\citeauthoryear{Grandi et al.}{2003}]{grandi03} Grandi, P., Guainazzi, M., Maraschi, L., Morganti, R., Fusco-Femiano, R., Fiocchi, M., Ballo, L., Tavecchio, F., 2003, ApJ, 586, 123

\bibitem[\protect\citeauthoryear{Groves et al.}{2004}]{groves04} Groves, B. A., Dopita, M. A., Sutherland, R. S., 2004, ApJS, 153, 75

\bibitem[\protect\citeauthoryear{Groves et al.}{2006}]{groves06} Groves, B. A., Heckman, T. M., Kauffmann, G., 2006, MNRAS, 371, 1559

\bibitem[\protect\citeauthoryear{Halpern \& Eracleous}{1994}]{halera94} Halpern, J. P., Eracleous, M., 2010, ApJ, 433, 17

\bibitem[\protect\citeauthoryear{Hardcastle \& Croston}{2005}]{hardcastle05} Hardcastle, M. J., Croston, J. H., 2005, MNRAS, 363, 649

\bibitem[\protect\citeauthoryear{Hardcastle et al.}{2015}]{hardcastle15} Hardcastle, M. J., Lenc, E., Birkinshaw, M., Croston, J. H.;, Goodger, J. L., Marshall, H. L., Perlman, E. S., Siemiginowska, A., Stawarz, L., Worrall, D. M., 2016, MNRAS, 455, 3526

\bibitem[\protect\citeauthoryear{Inskip et al.}{2010}]{inskip10} Inskip, K. J., Tadhunter, C. N., Morganti, R., Holt, J., Ramos Almeida, C., Dicken, D., 2010, MNRAS, 407, 1739

\bibitem[\protect\citeauthoryear{Lauberts}{1982}]{lauberts82} Lauberts, A., 1982, ESO/Uppsala survey of the ESO(B) atlas

\bibitem[\protect\citeauthoryear{Lauberts \& Valentijn}{1989}]{lauberts89} Lauberts, A., Valentijn, E. A., 1989, The surface photometry catalogue of the ESO-Uppsala galaxies 

\bibitem[\protect\citeauthoryear{Lena et al.}{2015}]{lena15} Lena, D., Robinson, A., Storchi-Bergman, T., Schnorr-Müller, A., Seelig, T., Riffel, R. A., Nagar, N. M., Couto, G. S., Shadler, L., 2015, ApJ, 806, 84

\bibitem[\protect\citeauthoryear{Lewis et al.}{2010}]{lewis10} Lewis, K. T., Eracleous, M., Storchi-Bergmann, T., 2010, ApJS, 187, 416

\bibitem[\protect\citeauthoryear{Lin et al.}{2013}]{lin13} Lin, L. H., Wang, H. H., Hsieh, P. Y., Taam, R. E., Yang, C. C., Yen, D. C. C., 2013, ApJ, 771, 8

\bibitem[\protect\citeauthoryear{Loveday}{1996}]{loveday96} Loveday, J., 1996, MNRAS, 1025, 1048

\bibitem[\protect\citeauthoryear{Lucy}{1974}]{lucy74} Lucy, L. B., 1974, AJ, 79, 745

\bibitem[\protect\citeauthoryear{Kewley et al.}{2006}]{kewley06} Kewley, L. J., Groves, B., Kauffmann, G., Heckman, T., 2006, MNRAS, 372, 961

\bibitem[\protect\citeauthoryear{Kewley et al.}{2013}]{kewley13} Kewley, L. J., Dopita, M. A., Leitherer, C., Davé, R., Yuan, T., Allen, M., Groves, B., Sutherland, R., 2013, ApJ, 774, 100

\bibitem[\protect\citeauthoryear{Kharb et al.}{2010}]{kharb10} Kharb, P., Lister, M. L., Cooper, N. J., 2010, ApJ, 710, 764

\bibitem[\protect\citeauthoryear{Kormendy \& Ho}{2013}]{korho13} Kormendy, J., Ho, L. C., 2013, ARA\&A, 51, 511

\bibitem[\protect\citeauthoryear{Marshall et al.}{2010}]{marshall10} Marshall, H. L., Hardcastle, M. J., Birkinshaw, M., Croston, J., Evans, D., Landt, H., Lenc, E., Massaro, F., Perlman, E. S., Schwartz, D. A., Siemiginowska, A., Stawarz, \L., Urry, C. M., Worrall, D. M., 2010, ApJ, 714, 213

\bibitem[\protect\citeauthoryear{Migliori et al.}{2007}]{miglioli07} Migliori, G., Grandi, P., Palumbo, G. G. C., Brunetti, G., Stanghellini, C., 2007, ApJ, 668, 203

\bibitem[\protect\citeauthoryear{Morganti et al.}{2013}]{morganti13} Morganti, R., Fogasy, J., Paragi, Z., Oosterloo, T., Orienti, M., 2013, Science, 341, 1082

\bibitem[\protect\citeauthoryear{Nesvadba et al.}{2009}]{nesvadba09} Nesvadba, N. P. H., Neri, R., De Breuck, C., Lehnert, M. D., Downes, D., Walter, F., Omont, A., Boulanger, F., Seymour, N., 2009, MNRAS, 395, 16

\bibitem[\protect\citeauthoryear{Oosterloo \& Morganti}{2005}]{oosmor05} Oosterloo, T. A., Morganti, R., 2005, A\&A, 429, 469

\bibitem[\protect\citeauthoryear{Osterbrock \& Ferland}{2006}]{ostfer06} Osterbrock, D. E., Ferland, G. J., 1989, Astrophysics of Gaseous Nebulae and Active Galactic Nuclei, 2nd. ed., University Science Books, California

\bibitem[\protect\citeauthoryear{Perley et al.}{1997}]{perley97} Perley, R. A., Roser, H. J., Meisenheimer, K., 1997, A\&A, 328, 12

\bibitem[\protect\citeauthoryear{Peterson}{1997}]{peterson97} Peterson, B. M., 1997, An Introduction to Active Galactic Nuclei, Cambridge University Press, Cambridge

\bibitem[\protect\citeauthoryear{Ramos Almeida et al.}{2011}]{ramos11} Ramos Almeida, C., Tadhunter, C. N., Inskip, K. J., Morganti, R., Holt, J., Dicken, D., 2011, MNRAS, 410, 1550

\bibitem[\protect\citeauthoryear{Reynaldi \& Feinstein}{2013}]{reyfei13} Reynaldi, V., Feinstein, C., 2013, MNRAS, 435, 1350

\bibitem[\protect\citeauthoryear{Reynolds et al.}{2002}]{reynolds02} Reynolds, C. S., Heinz, S., Begelman, M. C., 2002, MNRAS, 332, 271

\bibitem[\protect\citeauthoryear{Richardson}{1972}]{richardson72} Richardson, W. H., 1972, JOSA, 62, 55

\bibitem[\protect\citeauthoryear{Riffel}{2010}]{riffel10} Riffel, R. A., 2010, Ap\& SS, 327, 239


\bibitem[\protect\citeauthoryear{Rosario et al.}{2010}]{rosario10a} Rosario, D. J., Shields, G. A., Taylor, G. B., Salviander, S., Smith, K. L., 2010, ApJ, 716, 131

\bibitem[\protect\citeauthoryear{Santoro et al.}{2014}]{santoro14} Santoro, F., Oonk, J. B. R., Morganti, R., Oosterloo, T., 2015, A\&A, 574, 89

\bibitem[\protect\citeauthoryear{Schnorr-M{\"u}ller et al.}{2011}]{schnorr11} Schnorr-M{\"u}ller, A., Storchi-Bergmann, T., Riffel, R. A., Ferrari, F., Steiner, J. E., Axon, D. J., Robinson, A., 2011, MNRAS, 413, 149

\bibitem[\protect\citeauthoryear{Schnorr-M{\"u}ller et al.}{2014}]{schnorr14} Schnorr-M{\"u}ller, A., Storchi-Bergmann, T., Nagar, N. M., Robinson, A., Lena, D., Riffel, R. A., Couto, G. S., 2014, MNRAS, 437, 1708

\bibitem[\protect\citeauthoryear{Silverman et al.}{2008}]{silverman08} Silverman, J. D., Green, P. J., Barkhouse, W. A., Kim, D. -W., Kim, M., Wilkes, B. J., Cameron, R. A., Hasinger, G., Jannuzi, B. T., Smith, M. G., Smith, P. S., Tananbaum, H., 2008, ApJ, 679, 118

\bibitem[\protect\citeauthoryear{Simkin et al.}{1999}]{simkin99} 
Simkin, S. M., Sadler, E. M., Sault, R., Tingay, S. J., Callcut, J., 1999, ApJS, 123, 447

\bibitem[\protect\citeauthoryear{Steiner et al.}{2009}]{steiner09} Steiner, J. E., Menezes, R. B., Ricci, T. V., Oliveira, A. S., 2009, MNRAS, 395, 64

\bibitem[\protect\citeauthoryear{Storchi-Bergmann et al.}{1989}]{storchi89} Storchi-Bergmann, T., Pastoriza, M. G., 1989, ApJ, 347, 195

\bibitem[\protect\citeauthoryear{Storchi-Bergmann et al.}{1998}]{storchi98} Storchi-Bergmann, T., Schmitt, H. R., Calzetti, D., Kinney, A. L., 2009, AJ, 115, 909

\bibitem[\protect\citeauthoryear{Storchi-Bergmann et al.}{2010}]{storchi10} Storchi-Bergmann, T., Lopes, R. D. S., McGregor, P. J., Riffel, R. A., Beck, T., Martini, P., 2010, MNRAS, 402, 819

\bibitem[\protect\citeauthoryear{Tingay et al.}{2000}]{tingay00} Tingay, S. J., Jauncey, D. L., Reynolds, J. E., Tzioumis, A. K., McCulloch, P. M., Ellingsen, S. P., Costa, M. E., Lovell, J. E. J., Preston, R. A., Simkin, S. M., 2000, AJ, 119, 1695

\bibitem[\protect\citeauthoryear{Tingay et al.}{2008}]{tingay08} Tingay, S. J., Lenc, E., Brunetti, G., Bondi, M., 2008, AJ, 136, 2473

\bibitem[\protect\citeauthoryear{Tremblay et al.}{2009}]{tremblay09} Tremblay, G. R., Chiaberge, M., Sparks, W. B., Baum, S. A., Allen, M. G., Axon, D. J., Capetti, A., Floyd, D. J. E., Macchetto, F. D., Miley, G. K., Noel-Storr, J., O'Dea, C. P., Perlman, E. S., Quillen, A. C., 2009, ApJS, 183, 27

\bibitem[\protect\citeauthoryear{Tremonti et al.}{2004}]{tremonti04} Tremonti, C. A., Heckman, T. M., Kauffmann, G., Brinchmann, J., Charlot, S., White, S. D. M., Seibert, M., Peng, E. W., Schlegel, D. J., Uomoto, A., Fukugita, M., Brinkmann, J., 2004, ApJ, 613, 898

\bibitem[\protect\citeauthoryear{Urry \& Padovani}{1995}]{urry95} Urry, C. M., Padovani, P., 1995, PASP, 107, 803

\bibitem[\protect\citeauthoryear{Villar-Martin et al.}{1998}]{villar98} Villar-Martin, M., Tadhunter, C., Morganti, R., Clark, N., Killeen, N., Axon, D., 1998, A\&A, 332, 479

\bibitem[\protect\citeauthoryear{Wagner \& Bicknell}{2011}]{wagbic11} Wagner, A. Y., Bicknell, G. V., 2011, ApJ, 748, 29

\bibitem[\protect\citeauthoryear{Wilson et al.}{2001}]{wilson01} Wilson, A. S., Young, A. J., Shopbell, P. L., 2001, ApJ, 547, 740

\end{thebibliography}
\end{document}